\documentclass[12pt,indent]{article}
\usepackage{a4,latexsym,amsmath,amssymb}
\usepackage[latin1]{inputenc}
\usepackage{float}
\usepackage{natbib}
\usepackage{graphics}
\usepackage[english]{babel}
\usepackage{tabularx}
\usepackage{lscape}
\usepackage{multirow}
\usepackage{rotating}
\usepackage{dsfont}
\usepackage{caption}
\usepackage{subcaption}
\usepackage{bbm}
\usepackage[table]{xcolor}
\usepackage{booktabs}
\usepackage{calc}
\usepackage{ifthen}
\usepackage{url}
\usepackage{colortbl}      
\usepackage{tikz}

\usepackage{blindtext}
\usepackage{scrextend}
\usepackage{mathtools}
\usepackage{verbatim}

\usetikzlibrary{shapes,snakes,matrix}

\newenvironment{tabularsmall}
{ \footnotesize \sffamily \tabular } {
\endtabular
\normalfont }

\newcommand{\blanco}[1]{}

\def\d{\displaystyle}

\usepackage{natbib}

\usepackage{geometry}
\usepackage{setspace}
\usepackage{tikz}
\usepackage[]{graphicx}
\usepackage[]{color}
\makeatletter
\def\maxwidth{ %
  \ifdim\Gin@nat@width>\linewidth
    \linewidth
  \else
    \Gin@nat@width
  \fi
}
\makeatother

\definecolor{fgcolor}{rgb}{0.345, 0.345, 0.345}

\usepackage{framed}
\makeatletter
 {\par\unskip\endMakeFramed%
 \at@end@of@kframe}
\makeatother

\definecolor{shadecolor}{rgb}{.97, .97, .97}
\definecolor{messagecolor}{rgb}{0, 0, 0}
\definecolor{warningcolor}{rgb}{1, 0, 1}
\definecolor{errorcolor}{rgb}{1, 0, 0}

\usepackage{alltt}
\IfFileExists{upquote.sty}{\usepackage{upquote}}{}

\newtheorem{theorem}{Proposition}[section]

\begin{document}
\bibliographystyle{chicago}
\sloppy

\makeatletter
\renewcommand{\section}{\@startsection{section}{1}{\z@}%
        {-3.5ex \@plus -1ex \@minus -.2ex}%
        {1.5ex \@plus.2ex}%
        {\reset@font\Large\sffamily}}
\renewcommand{\subsection}{\@startsection{subsection}{1}{\z@}%
        {-3.25ex \@plus -1ex \@minus -.2ex}%
        {1.1ex \@plus.2ex}%
        {\reset@font\large\sffamily\flushleft}}
\renewcommand{\subsubsection}{\@startsection{subsubsection}{1}{\z@}%
        {-3.25ex \@plus -1ex \@minus -.2ex}%
        {1.1ex \@plus.2ex}%
        {\reset@font\normalsize\sffamily\flushleft}}
\makeatother



\newsavebox{\tempbox}
\newlength{\linelength}
\setlength{\linelength}{\linewidth-10mm} \makeatletter
\renewcommand{\@makecaption}[2]
{
  \renewcommand{\baselinestretch}{1.1} \normalsize\small
  \vspace{5mm}
  \sbox{\tempbox}{#1: #2}
  \ifthenelse{\lengthtest{\wd\tempbox>\linelength}}
  {\noindent\hspace*{4mm}\parbox{\linewidth-10mm}{\sc#1: \sl#2\par}}
  {\begin{center}\sc#1: \sl#2\par\end{center}}
}



\def\R{\mathchoice{ \hbox{${\rm I}\!{\rm R}$} }
                   { \hbox{${\rm I}\!{\rm R}$} }
                   { \hbox{$ \scriptstyle  {\rm I}\!{\rm R}$} }
                   { \hbox{$ \scriptscriptstyle  {\rm I}\!{\rm R}$} }  }

\def\N{\mathchoice{ \hbox{${\rm I}\!{\rm N}$} }
                   { \hbox{${\rm I}\!{\rm N}$} }
                   { \hbox{$ \scriptstyle  {\rm I}\!{\rm N}$} }
                   { \hbox{$ \scriptscriptstyle  {\rm I}\!{\rm N}$} }  }

\def\d{\displaystyle}\def\d{\displaystyle}

\title{On the Structure of Ordered Latent Trait Models }
 \author{Gerhard Tutz \\{\small Ludwig-Maximilians-Universit\"{a}t M\"{u}nchen}\\{\small Akademiestra{\ss}e 1, 80799 M\"{u}nchen}}


\maketitle

\begin{abstract} 
\noindent
Ordered item response models that are in common use  can be divided into three groups, cumulative, sequential and adjacent categories model.
The derivation and motivation of the models is typically based on the assumed presence of latent traits or underlying process models. In the construction frequently  binary models play an important role. The objective of this paper is to give motivations for the models and to clarify the role of the  binary models for the various types of ordinal models.  It is investigated which binary models are included in an ordinal model but also how the models can be constructed from a sequence of binary models. In all the models one finds a Guttman space structure, which has  previously been investigated in particular for the partial credit model. The consideration of the binary models adds to the interpretation of model parameters, which is helpful, in particular, in the case of the partial credit model, for which interpretation is less straightforward than for the other models. A specific topic that is addressed is the ordering of thresholds in the partial credit model because for some researchers reversed ordering is an anomaly,  others disagree. It is argued that the ordering of thresholds is not a constitutive element of the partial credit model.


\end{abstract}


\section{Introduction}

Various latent trait models for ordered response data have been proposed in the literature, for an overview see, for example, \citet{VanderLind2016}.  One can distinguish between three basic types of models, cumulative models, sequential models and adjacent categories models. While for the first two models the interpretation of the latent trait is straightforward, there is still ongoing discussion on the interpretation of the parameters in the adjacent categories approach, see, for example, \citet{adams2012rasch} or \citet{andrich2015problem}.  In the latter article the problems with the step  metaphor in polytomous models, which is a source of misinterpretation,   are extensively discussed.
There has also been some discussion of the ordering of thresholds in adjacent categories models. \citet{andrich2013expanded}  strongly criticized the findings of \citet{adams2012rasch} and argued that reverse order of thresholds is an anomaly. One motivation for this paper was to clarify some of the derivations of \citet{andrich2013expanded}. In particular it is argued that the ordering of thresholds is not what makes the adjacent categories model an ordered latent trait model. Moreover, a simpler form of the Guttman structure in the partial credit model is derived.

Ordered latent trait models are typically motivated by latent random variables because the latent process is useful to clarify the meaning of parameters.  By characterizing the process that generates  the response  in one of the ordered categories  one obtains an understanding of the meaning of the parameters. A specific property of all the ordered models considered here is that they are  linked to binary models, typically binary Rasch models.
One of the objectives of the present paper is to investigate the nature of this link. In the investigation it is often useful to distinguish two different views on the link to binary models.  One can start from an ordered latent trait model and identify which binary models are implied. Alternatively, one can start from binary models and derive from them an ordered response model. Both views can contribute to the interpretation of the model parameters. 

As a preview on the structure of models and the involved binary variables 
Table \ref{tab:ORDINAL} shows the three types of
dichotomizations that determine the models. It is  assumed that the ordinal response $Y$ takes values from $\{0,1,\dots,k\}$. The variables $Y^{(r)}$ denote the binary variables that are contained in the model. Brackets denote which
response categories are used when dichotomizing and the "$|$"
determines the (conditional) split. The table shows which binary models are contained in ordered latent trait models but can also be used to construct models from binary variables.

\begin{table}[h!]
\small
\caption{Types of ordinal models for response $Y  \in \{0,1,\dots,k\}$ and dichotomous variables $Y^{(r)}$
that are contained.} \label{tab:ORDINAL}
\begin{center}
\fbox{%
\parbox[t]{\linewidth-3\fboxsep}{%
\centerline{}
\centerline{\textsf{Cumulative-Type Model, Dichotomization into
Groups}}
\[
\begin{array}{cc}
[0,\dots, r-1|r,\dots, k]\quad & \quad Y^{(r)}= \left\{
\begin{array}{ll}
1 & Y\in\{r,\dots, k\}\\
0 & Y\in\{0,\dots, r-1\}\\
\end{array}
\right.
\end{array}
\]
\centerline{}
\centerline{\textsf{Sequential-Type Model, Dichotomization Given
$Y\geq r$}}
\[
\begin{array}{cc}
0,\dots, [r-1|r,\dots, k]\quad & \quad Y^{(r)}= \left\{
\begin{array}{ll}
1 & Y \geq r\\
0 & Y < r\\
\end{array}
\right.
\end{array}
\mbox{given}\,\,\, Y\geq r-1
\]
\centerline{}
\centerline{\textsf{Adjacent-Type Model, Dichotomization Given
$Y\in\{r, r+1\}$}}
\[
\begin{array}{cc}
0,\dots, [r-1|r],\dots, k\quad & \quad Y^{(r)}= \left\{
\begin{array}{ll}
1 & Y=r\\
0 & Y < r\\
\end{array}
\right.
\end{array}
\mbox{given}\,\,\, Y\in\{r-1, r\}
\]
}}
\end{center}
\end{table}

We start with cumulative models  and then consider sequential models and adjacent categories models. The latter is investigated more closely because the binary variables involved are much  harder to investigate than for the other models.

\section{Cumulative or Graded Response Models}\label{sec:Cum}

We first consider cumulative models, for which interpretation of parameters is most easily obtained by deriving the model from latent variables. In the following let $Y_{pi} \in \{0,1,\dots,k\}$, $p=1,\dots,P$, $i=1,\dots,I$, denote the ordinal response of person $p$ on item $i$.

\begin{figure}[h!]
\begin{center}
\includegraphics[width=10cm]{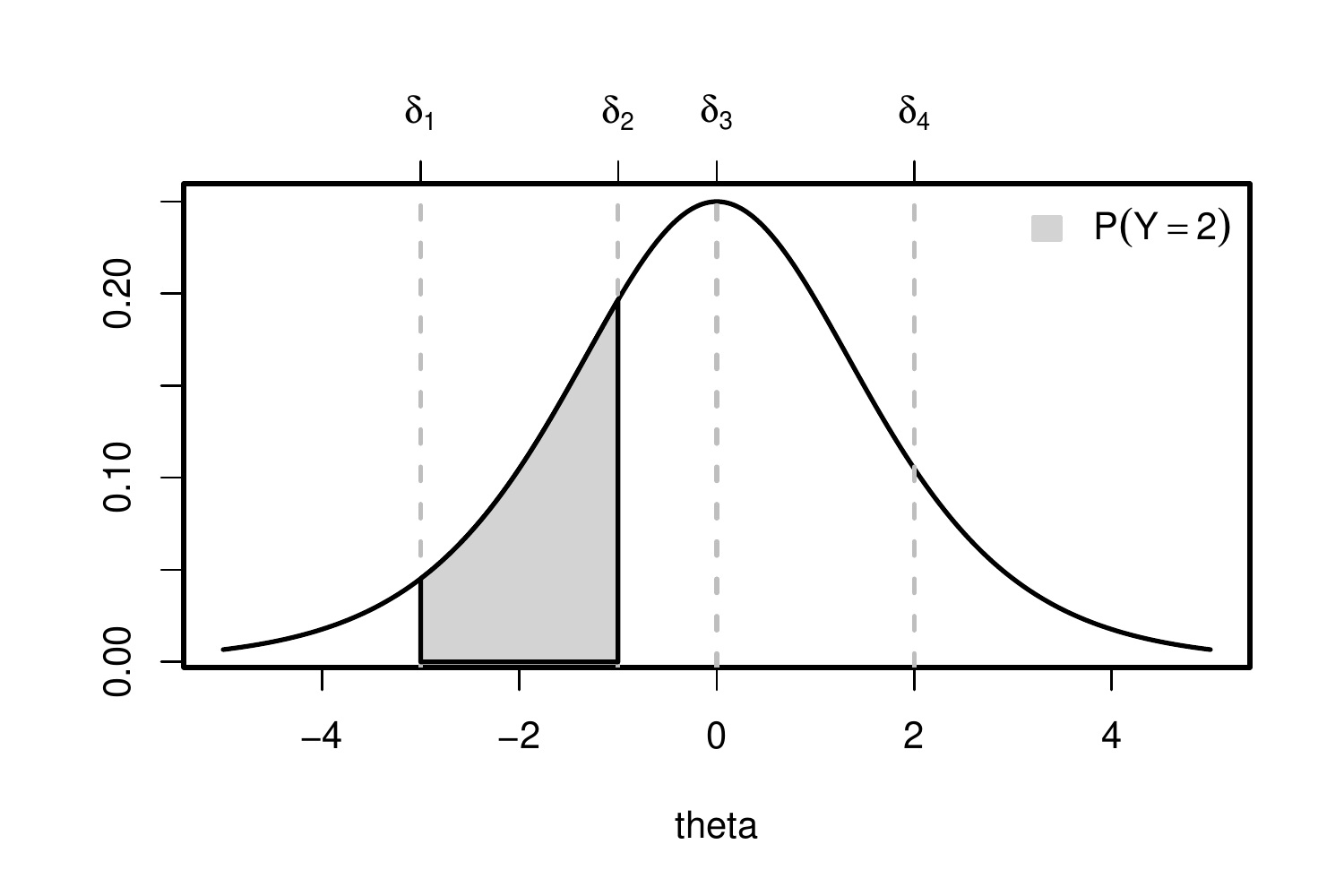}
\caption{Density of the underlying latent trait; the shaded area between thresholds $\delta_1$ and $\delta_2$ represents the probability of observing category two, $P(Y=2)$.}\label{fig:cutoff}
\end{center}
\end{figure}

\subsection{Model Derivation from Latent Variables}

The class of \textit{cumulative} models can be motivated by a latent variable  $\tilde Y_{pi}=\theta_p+\varepsilon_{pi}$, where $\theta_p$ is a person parameter and $\varepsilon_{pi}$ is a noise variable with symmetric continuous distribution function $F(.)$. The category boundaries approach assumes that the observed response is a coarser version of the latent variable $\tilde Y_{pi}$. More concrete, if person $p$ meets item $i$ one assumes
\[
Y_{pi}=r\quad\Leftrightarrow\quad\delta_{ir}\leq \tilde Y_{pi}<\delta_{i,r+1},
\]
where $\delta_{i1}\le\dots\le\delta_{ik}$ are ordered item-specific thresholds.
Thus, one observes a response in category $r$ if the latent variable is between thresholds $\delta_{ir}$ and $\delta_{i,r+1}$. For an illustration see Figure \ref{fig:cutoff}, which shows the density of the latent variable if $\theta_p=0$. It is easy to derive the cumulative model 
\[
P(Y_{pi} \ge r)=F(\theta_p-\delta_{ir}), \quad r=1,\dots,k.
\]
For the logistic distribution function $F(\eta)=\exp(\eta)/(1+\exp(\eta))$ and $k=1$ one obtains the simple binary Rasch model $P(Y_{pi} =1)=\exp(\theta_p-\delta_{i })/(1+\exp(\theta_p-\delta_{i }))$. The general model with $k>1$  is equivalent to Samemija's \textit{graded response model} \citep{samejima1997graded, samejima2016graded}.

It is implicitly assumed that the person parameters $\theta_p$ are on the same latent scale as the 
item parameters $\delta_{i1},\dots,\delta_{ik}$.   The item parameters have a distinct interpretation as thresholds on the latent scale  and the person parameters represent the location of the persons on the latent scale.  In proficiency tests the person parameter  can be interpreted as ability and the item parameters as difficulties. Person $p$ has location $\theta_p$ on the latent ability scale, but ability is blurred by the noise variable $\varepsilon_{pi}$. Nevertheless, primarily  the location determines the observed probabilities. If $\theta_p$ is large the probability $P(Y_{pi}=k)$ will be large and the probability $P(Y_{pi}=1)$ will be small. For small $\theta_p$ the order is reversed.  For fixed $\theta_p$ the difference $\delta_{i,r+1}-\delta_{ir}$  determines if the probability $P(Y_{pi}=r)$ 
tends to be large or small. 


\subsection{ Binary Responses in the Cumulative Models }

A special feature of the model is that the binary Rasch model (for $F(.)$ logistic) holds if   categories are collapsed  into two groups. More specific, if one considers the grouped response $Y_{pi}^{(r)}=1$ if $Y_{pi} \in \{r,\dots,k\}$ and $Y_{pi}^{(r)}=0$, otherwise, one obtains for $Y_{pi}^{(r)}$ the binary Rasch model with predictor $\theta_p-\delta_{ir}$, 
\[
P(Y_{pi}^{(r)}=1)=F(\theta_p-\delta_{ir}).
\]
For the link between the binary variables and the response in $k+1$ categories one obtains
\[
Y_{pi}=r   \quad\Leftrightarrow\quad (Y_{pi}^{(1)}\dots,Y_{pi}^{(k)})=(1,\dots,1,0,\dots,0),
\]
where $Y_{pi}^{(r)}$ is the last of the sequence of binary variables with an value 1. All the sequences $(Y_{pi}^{(1)}\dots,Y_{pi}^{(k)})$ that result have the property that a sequence of ones is followed by a sequence of zeros, which reminds of a Guttman scale. Therefore,   random variables $Y_{pi}^{(1)}\dots,Y_{pi}^{(k)}$ with $Y_{pi}^{(r)} \in \{0,1\}$ and $Y_{pi}^{(r)} \ge Y_{pi}^{(r+1)}$  will be called Guttman variables (they have a Guttman structure), the corresponding Guttman space is defined by $\Omega^G=\{(y_1,\dots,y_k)|y_r \in \{0,1\}, y_r \ge y_s \text{ for all } r < s\}$.
The binary variable $Y_{pi}^{(r)}$ contains the information if  a response is larger than or equal to category $r$ or not. If $Y_{pi}^{(r)}=1$ one knows that the response is in category $r$ or in a higher one.


\subsection{Binary Models as Building Blocks}\label{sec:BinBuild}

The cumulative model can also be constructed from binary variables. 
Let the binary variables be defined as above by $Y_{pi}^{(r)}=1$ if $Y_{pi} \in \{r,\dots,k\}$ and $Y_{pi}^{(r)}=0$, otherwise.  If one assumes that the binary response models $P(Y_{pi}^{(r)}=1)=F(\theta_p-\delta_{ir})$ hold for $r=1,\dots,k$ one obtains immediately the cumulative model $P(Y_{pi}\ge r)=F(\theta_p-\delta_{ir})$. Thus, the binary variables can be considered the building blocks of the cumulative model. 

It should be noted that the binary variables used here are defined with reference to the ordinal response. Therefore, one implicitly assumes  that an ordinal response   is present. As   shown in the following section one can also construct the model  by starting with a specific set of linked binary models and then define an ordinal response.

\subsection*{Deriving the Model  Without  Using the Ordinal Response}

Cumulative models can also be constructed from binary response models without reference to an already given ordinal response. 
Consider a set of binary responses $Y_{pi}^{(r)}, r=1,\dots, k$. Let us assume for the binary variables:
\begin{itemize}
\item[] 
(a)  Binary models  $P(Y_{pi}^{(r)}=1)=F(\theta_p-\delta_{ir})$  hold for $r=1,\dots,k$.  
\item[] 
(b) The binary variables are Guttman variables, that is, $Y_{pi}^{(r)} \ge Y_{pi}^{(s)}$  holds  for $r <s$. 
\end{itemize}
As will be shown  one can construct an ordinal response model from these binary variables. This is possible since the assumption that the binary variables are Guttman variables implies an order.

\begin{theorem}
Let assumptions (a) and (b) hold. For the ordinal response variable $Y_{pi}$ defined by 
\[
Y_{pi}=Y_{pi}^{(1)}+\dots+Y_{pi}^{(k)}
\]
one obtains the cumulative model 
\[
P(Y_{pi} \ge r)=F(\theta_p-\delta_{ir})
\]
iff 
\begin{equation}\label{eq:cond2}
P(Y_{pi}^{(r)}=1) \ge P(Y_{pi}^{(s)}=1) \text{ for } r <s.
\end{equation} 
\end{theorem}

To illustrate why this holds let us consider the case of three categories $0,1,2$. Then, one has two binary variables $Y_{pi}^{(1)},Y_{pi}^{(2)}$. Let $\pi_r=P(Y_{pi}^{(r)}=1)$, $r=1,2$, denote the   probabilities of obtaining 1 in  the binary variables.
In the   contingency table given in Table \ref{tab:cum}  all the probabilities are determined by the marginal probabilities (which are fixed) and the assumption that one entry has probability zero, which follows from assumption (b).  

\begin{table}[h!]
\caption{Contingency tables for the  binary variables $( Y_{pi}^{(1)},  Y_{pi}^{(2)} )$  }\label{tab:cum}
\centering

  \begin{minipage}{0.5\textwidth}
\begin{tabular}{cc|cc|c}
                           & \multicolumn{1}{c}{}            & \multicolumn{2}{c}{$Y_{pi}^{(2)}$}&                 \\
                           & \multicolumn{1}{c}{} & 1 & \multicolumn{1}{c}{0}&         \\ \cline{3-4}
                           & 1                        & $\pi_2$  & $\pi_1-\pi_2$                          & $\pi_1$      \\
\raisebox{1.5ex}[-1.5ex]{$Y_{pi}^{(1)}$} & 0                        &
$0$ & $1-\pi_1$ & $1-\pi_1$      \\ \cline{3-4}
                           &  \multicolumn{1}{c}{}         & $\pi_2$      & \multicolumn{1}{c}{$1-\pi_2$}                          & $1$      \\
\end{tabular}
\normalsize
  \end{minipage}
\end{table}

This is a valid contingency table only if $ \pi_1 \ge \pi_2$, which corresponds to  the condition (\ref{eq:cond2}). For $Y_{pi}=Y_{pi}^{(1)}+Y_{pi}^{(2)}$  one obtains immediately the probabilities $P(Y_{pi}=r)$ as

\begin{table}[H]
\centering
\begin{tabular}{@{}cccc@{}}
\multicolumn{3}{c}{}\\
 $r$   &0                & 1   & 2\\\midrule
 $P(Y_{pi}=r)$    &$1-\pi_1$          &$\pi_1-\pi_2$ &$\pi_2$   \\
\bottomrule
\end{tabular}
\end{table}

\noindent 
and therefore $P(Y_{pi} \ge 2)=\pi_2$, $P(Y_{pi} \ge 1)=\pi_1$, and $P(Y_{pi} \ge 0)=1$. Since by assumption $\pi_r=F(\theta_p-\delta_{ir})$ one obtains the cumulative model.

Proposition 2.1 shows that a cumulative model can be constructed from a sequence of binary variables. For the marginal distributions of the binary variables   binary Rasch models (for $F(.)$ logistic) are assumed to hold with all of the models containing the parameter $\theta_p$. It means that the cumulative model does not only contain binary Rasch models but can also be built from them. The condition (b), that is, $Y_{pi}^{(r)} \ge Y_{pi}^{(s)}$ for $r<s$, means that the sequences of binary variables have the form $(1,\dots,1,0,\dots,0)$. The binary variables are not independent, and are, in particular, restricted by $P(Y_{pi}^{(r)}=1) \ge P(Y_{pi}^{(s)}=1)$ for $r<s$. The latter condition is equivalent to the ordering of the thresholds, $\delta_{ir} \le \delta_{is}$ for $r<s$. The binary variables also show that the parameters can be interpreted as the parameters of binary Rasch models, in particular the parameters $\delta_{ir}$ represent difficulty parameters if one has a proficiency test.

The derivation of the cumulative model from binary variables that are not defined by an ordinal response but yield an ordinal response might seem not to be crucial for the interpretation of models. However, it should be seen in connection with the motivation of the partial credit model, to be treated later.    \citet{andrich2013expanded} tried to motivate the   partial credit model by considering binary variables in a  
similar but nevertheless distinctly different way (see Section \ref{sec:BuildBlocks}).

Let us finally comment on the role of the Rasch model when deriving an ordinal response model from binary models. In the derivation it was assumed that the binary variables follow a Rasch model. However, one can drop this assumption and only postulate that $Y_{pi}^{(r)} \ge Y_{pi}^{(s)}$  holds  for $r <s$ (assumption (b)). Then one obtains for the ordinal response defined by $Y_{pi}=Y_{pi}^{(1)}+\dots+Y_{pi}^{(k)}$
the link $P(Y_{pi} \ge r)=P(Y_{pi}^{(r)}=1)$ provided $P(Y_{pi}^{(r)}=1) \ge P(Y_{pi}^{(s)}=1)$ holds for $r<s$. Thus the construction of a cumulative model does not have to assume a specific model for the binary variables. If, however, one specifies the marginals by  a binary model, that is, $P(Y_{pi}^{(r)}=1)=F(\theta_p-\delta_{ir})$ one obtains the corresponding cumulative latent trait model.  

In summary, it is seen that the existence of binary variables, for which $Y_{pi}^{(r)} \ge Y_{pi}^{(r+1)}$ and $P(Y_{pi}^{(r)}=1) \ge P(Y_{pi}^{(r+1)}=1)$  hold    is a necessary and sufficient condition for the cumulative model.

\begin{figure}[ht]
\begin{tikzpicture}[level distance=20mm, thick, sibling distance=45mm]
\tikzstyle{every node}=[circle, inner sep=3pt]
\tikzstyle{level 5}=[level distance = 8mm] \node [circle,
draw]{Start} child {node[circle, draw]{$Y_{pi}=0$} edge from parent
node[left]{$Y_{pi}^{(1)}=0$}
    child[grow=down, level distance=20mm]{node{$$} edge from parent [draw=none]{} }}
child {node [circle, draw]{$\geq 1$}
    child {  node [circle, draw]{$Y_{pi}=1$} edge from parent node[left]{$Y_{pi}^{(2)} = 0$}
        child[grow=down, level distance=20mm]{node {$$} edge from parent[draw=none]} }
    child {node [circle, draw]{$\geq 2$}
        child {node [circle, draw]{$Y_{pi}=2$} edge from parent node[left]{$Y_{pi}^{(3)} = 0$}
            child[grow=down, level distance=20mm]{node {$$} edge from parent[draw=none]}}
        child {node[circle, draw] {$\geq 3$}
            child {node [circle, draw]{$Y_{pi}=3$} edge from parent node[left]{$Y_{pi}^{(4)} = 0$}
                child[grow=down, level distance=20mm]{node {$$} edge from parent[draw=none]}}
            child {node [circle, draw]{$Y_{pi}=4$}
                child[level distance=11mm]{node {$$}edge from parent[draw=none]}}
        }
    }
};
\end{tikzpicture}
\caption{Steps for item with categories $0,1,2,3,4$} \label{bild.baum}
\end{figure}
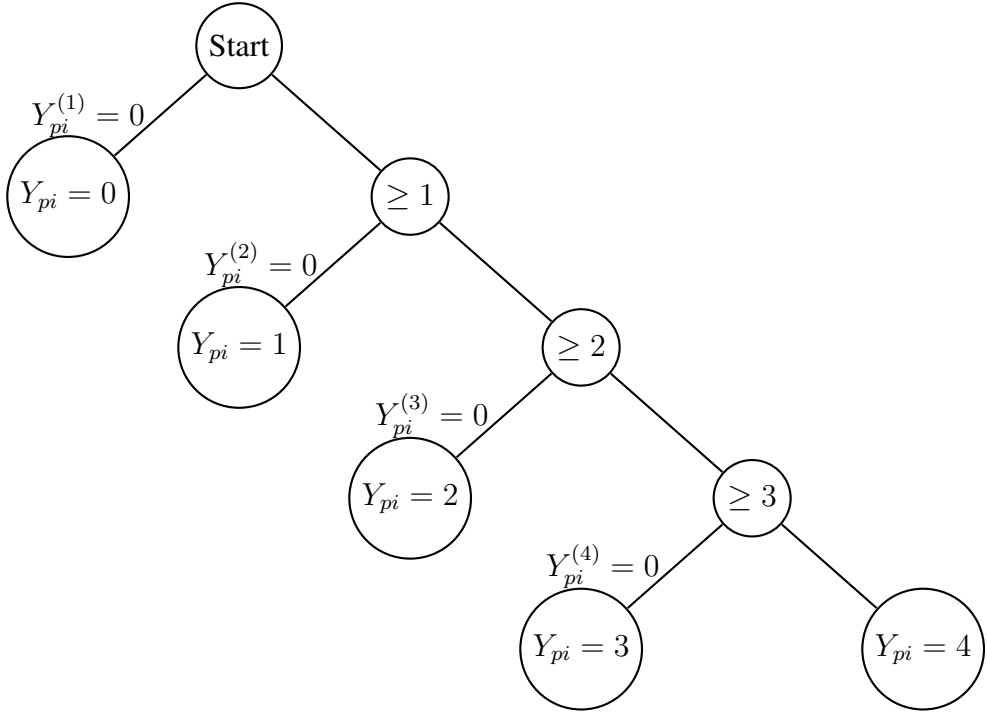

\section {Sequential Models}
We start with a traditional derivation of the model and then consider the binary variables that are linked to the model.

\subsection {Derivation of the Model from Consecutive Steps}

The sequential model is typically used in proficiency tests where a person tries to solve an item in several steps with the scores representing the levels of the solution. A classical example considered by    
\citet{Masters:82} is the item $\sqrt{9.0/0.3-5}=?$. Three levels of performance may be distinguished: No sub-problem solved (level 0), 9.0/0.3 = 30 solved (level 1), $30 ? 5 = 25$ solved (level 2), $\sqrt{25} = 5$ (level 3). The main feature of the sequential or step model is that the stepwise manner of solving an item is exploited. 

In each of the consecutive steps a  binary response model applies. Let $Y_{pi}^{(r)}$ denote the binary response in the $r$-th step. In the first step the respondent tries to master the transition from level 0 to level 1. The corresponding model is 
\[
P(Y_{pi}^{(1)}=1)=F(\theta_p-\delta_{i1}).
\]
If the person does not master the first step, the process stops and one observes $Y_{pi}=0$. Otherwise,   $Y_{pi} \ge 1$ and the person tries to solve the next step. In general, the $r$-th step, which is only tried if the previous steps were successful,  is determined by 
\[
P(Y_{pi}^{(r)}=1|Y_{pi}\ge r-1)=F(\theta_p-\delta_{ir}), \quad r=1,\dots,k.
\]
If the person does not master the  step, the process stops and one observes $Y_{pi}=r-1$. Otherwise,   $Y_{pi} \ge r$ and the person tries to solve the next step.
It is straightforward to derive the sequential model 
\begin{equation}\label{eq:seq}
P(Y_{pi}\ge r|Y_{pi}\ge r-1)=F(\theta_p-\delta_{ir}), \quad r=1,\dots,k.
\end{equation}
The process model is illustrated in Figure \ref{bild.baum} for response categories $\{0,\dots,4\}$.

The important feature of the model is that the transition from one level to the next is modelled \textit{conditionally} (given the previous level has been reached). For each conditional step the 
binary model could also be motivated by  latent random variables $\tilde Y_{pi}^{(r)}=\theta_p+\varepsilon_{pi}$, where  $\varepsilon_{pi}$ is a noise variable with symmetric continuous distribution function $F(.)$. Given the $r$-th step is reached one has $Y_{pi}^{(r)}=1$, or  equivalently $Y_{pi}\ge r$, if $\tilde Y_{pi}^{(r)}>\delta_{ir}$, and $Y_{pi}^{(r)}=0$ ($Y_{pi}= r$) otherwise. Thus, in proficiency tests it is straightforward to interpret the person parameter as the ability and $\delta_{ir}$ as the difficulty of the $r$-th step. The parameter $\delta_{ir}$ can   be seen as a threshold that has to be mastered in the 
$r$-th step. No ordering of the difficulties/thresholds $\delta_{i1}, \dots, \delta_{ik}$ is assumed. For example, the third step can be easier than the second step ($\delta_{i3}<\delta_{i2}$). However, the person has to solve all the previous steps before being able to  try to master the third step. The ordering of the measurement is due to the sequential nature of the model.  

For the probabilities one derives 
\begin{align*}
P(Y_{pi}=r) 
=(1-F(\theta_p-\delta_{i,r+1}))\prod_{j=1}^{r}F(\theta_p-\delta_{ij}), \quad r=0,\dots,k,
\end{align*}
where $\prod_{j=1}^{0} \equiv 1$. The probabilities also reflect the sequential nature of $r$ successful steps with probabilities $F(\theta_p-\delta_{ij})$ followed by one unsuccessful step. In the $r$th step the   probability  of failing is $1-F(\theta_p-\delta_{i,r+1})$.

The cumulative model has been motivated by a latent variable and then binary models that can be used to construct the model have been considered. The motivation for the sequential model given here is different. It starts with the binary variables, which represent the conditional transitions. Thus, the model itself is constructed by using  binary models.

It is straightforward to show that
the  binary variables, which determine the transitions, can also be derived from the sequential model (\ref{eq:seq}).  One assumes that the sequential model holds and uses the dichotomization into groups $\{r-1\}$ and $\{r,\dots,k\}$ \textit{given $Y_{pi} \ge r-1$} to define  

\[
Y_{pi}^{(r)}=\left\{
\begin{array}{ll}
1&Y_{pi} \geq r \text{ given } Y_{pi} \ge r-1\\
0&Y_{pi} < r \text{ given } Y_{pi} \ge r-1.
\end{array}
\right.
\]
These are exactly the conditional variables given in Table \ref{tab:ORDINAL}.

The first version of this model was considered by \citet{molenaar1983item}. \citet{Tutz:90b} considered a parametric model  under the name \textit{sequential model}, \citet{verhelst1997steps}  called the corresponding model \textit{step model}. The model is strongly related to the  class of continuation ratio models, see, for example, \citet{Agresti:2013}, \citet{TutzBook2011}. Measurement properties of the model were investigated by \citet{hemker2001measurement}.

\subsection {Derivation from Binary Variables}

In the previous section the resulting ordinal response was explained by consecutive steps that model the transitions between categories.  However, as for the cumulative model one can also start with Guttman variables and then construct an ordinal response without using that an ordinal experiment has been executed.

Let us again consider a set of binary variables $Y_{pi}^{(1)}\dots,Y_{pi}^{(k)}$, $Y_{pi}^{(r)} \in \{0,1\}$,  and assume 
\begin{itemize}
\item[] 
(a)  A conditional binary model  $P(Y_{pi}^{(r)}=1|Y_{pi}^{(r-1)}=1)=F(\theta_p-\delta_{ir})$  holds for $r=2,\dots,k$.   
\item[] 
(b) The binary variables are Guttman variables, that is, $Y_{pi}^{(r)} \ge Y_{pi}^{(s)}$  holds  for $r <s$. 
\end{itemize} 

An ordinal response model can be constructed in the following way.

\begin{theorem}\label{th:Sequ}
Let assumptions (a) and (b) hold and the ordinal response variable $Y_{pi}$ be defined by 
\[
Y_{pi}=r   \quad\Leftrightarrow\quad (Y_{pi}^{(r)},Y_{pi}^{(r+1)} )=(1,0),
\]
where, for completeness, $Y_{pi}^{(0)}$ is defined by $Y_{pi}^{(0)}\equiv 1$.
Then,  for $Y_{pi}$  the sequential model holds, that is, 
\[
P(Y_{pi} \ge r|Y_{pi} \ge r-1)=F(\theta_p-\delta_{ir}).
\]
 
\end{theorem}

It is not hard to see that Proposition \ref{th:Sequ} holds. It follows from assumption (a) that all vectors show a Guttman structure, that is,  they have the form  $(1,\dots, 1,0\dots,0)$. The definition of $Y_{pi}$ is equivalent to
\[
Y_{pi}=r   \quad\Leftrightarrow\quad (Y_{pi}^{(1)},\dots, Y_{pi}^{(r+1)})=(1,\dots, 1,0).
\]
That means $Y_{pi}=r$ if $Y_{pi}^{(r)}$ is the last in the sequence of Bernoullis $Y_{pi}^{(1)},Y_{pi}^{(2)},\dots$, that has response 1.
Since the binary variables have a Guttman structure,  one can also use the sum of the binary variables, $Y_{pi}=\sum_{j=1}^{k}Y_{pi}^{(j)}$ to define the ordinal response. One uses the general result   
\begin{align*} 
&P(Y_{pi}^{(0)}=1,Y_{pi}^{(1)}=r_1,\dots,Y_{pi}^{(k)}=r_k)\\&
=P(Y_{pi}^{(k)}=r_k|Y_{pi}^{(k-1)}=r_{k-1})\dots P(Y_{pi}^{(1)}=r_1|Y_{pi}^{(0)}=1)P(Y_{pi}^{(0)}=1), 
\end{align*}
the property $P(Y_{pi}^{(0)}=1)=1$, and $P(Y_{pi}^{(r)}=0|Y_{pi}^{(r-1)}=0)=1$,  which follows  from assumption (b), to obtain the probabilities
\begin{align*}
P(Y_{pi} = 0)&= P((Y_{pi}^{(1)},\dots,Y_{pi}^{(k)})=(0,\dots,0))=1-F(\theta_p-\delta_{i1}),\\
P(Y_{pi} = r)&= P(Y_{pi}^{(r+1)}=0|Y_{pi}^{(r)}=1)\prod_{j=1}^{r} P(Y_{pi}^{(j)}=1|Y_{pi}^{(j-1)}=1)\\
&= (1-F(\theta_p-\delta_{i,r+1}))\prod_{j=1}^{r} F(\theta_p-\delta_{ij}), r >0,
\end{align*}
which are the response probabilities of the sequential model.

The sequence of observations $(1,\dots, 1,0)$ is at the heart of the sequential model. It directly reflects the successive transitions of $r-1$
successful steps followed by an unsuccessful step. The sequence $Y_{pi}^{(1)},\dots,Y_{pi}^{(k)}$ can also be seen  a Markov chain  that stops if the absorbing state 0 is reached.

It is interesting to compare the construction of the cumulative and the sequential models from binary variables. In both derivations the main assumption is the Guttman structure. No further assumption is needed when defining an ordinal response that yields the sequential model, whereas one needs the probabilistic monotonicity assumption  $P(Y_{pi}^{(r)}=1) \ge P(Y_{pi}^{(r+1)}=1)$  to construct a cumulative model. Of course, the role of the Rasch model differs. In the cumulative model the marginal distribution of the Guttman vectors follows a Rasch model, while the Rasch model is assumed for the conditional transitions in the sequential model. 
It should be noted that for both models the ordinal variables $Y_{pi}$  as well as the involved binary variables are observable.

\section{Adjacent Categories: The Partial Credit Model}\label{sec:PCM}

In contrast to the cumulative and sequential model, for which interpretation of the parameters is straightforward, the interpretation in the adjacent categories model is much more  challenging. In the following, we start with a brief description of the partial credit model and consider the motivation of the model later.

\subsection{ The Partial Credit Model}\label{sec:PCM}
The most prominent member of the family of adjacent categories models is the partial credit model (PCM), which was proposed by  \citet{Masters:82}, see also \citet{MasWri:84} and extensions given by \citet{glas1989extensions}, \citet{von2004partially}.
The basic PCM  assumes for the probabilities
\[
P(Y_{pi}=r)= \frac{\exp(\sum_{l=1}^{r}(\theta_p-\delta_{il}))}{\sum_{s=0}^{k}\exp(\sum_{l=1}^{s}(\theta_p-\delta_{il}))}, \quad r=1,\dots,k,
\]
where $\theta_p$ is the person parameter and $\delta_{i1},\dots,\delta_{ik}$ are the item parameters of item $i$.
For notational convenience   the definition of the model implicitly uses $\sum_{k=1}^{0}\theta_p-\delta_{ik}=0$.
With this convention an alternative form is given by
\[
P(Y_{pi}=r)= \frac{\exp(r\theta_p-\sum_{l=1}^{r}\delta_{il})}{\sum_{s=0}^{k}\exp(\sum_{l=1}^{s}(\theta_p-\delta_{il}))}.
\]
The defining property of the partial credit model is seen if one considers adjacent categories. For the comparison of two adjacent categories one obtains immediately 
\[
\log \left(\frac {P(Y_{pi}=r)}{P(Y_{pi}=r-1)}\right)= \theta_p-\delta_{ir},\quad r=1,\dots,k.
\]
However, it is often more useful to consider the representation
\begin{equation}\label{eq:PCMDef}
\frac {P(Y_{pi}=r)}{P(Y_{pi}=r-1)+P(Y_{pi}=r)}= F(\theta_p-\delta_{ir}),\quad r=1,\dots,k,
\end{equation}
where $F(.)$ is the logistic function. It shows that the model is \textit{locally (given response categories $r-1$, $r$)}  a binary Rasch model with person parameter $\theta_p$ and item difficulty $\delta_{ir}$. The general form of the model is
\begin{equation}\label{eq:PCM}
{P(Y_{pi}=r|Y_{pi} \in \{r-1,r\})}= F(\theta_p-\delta_{ir}),\quad r=1,\dots,k,
\end{equation}
where $F(.)$ is a continuous cumulative distribution function. If $F(.)$ is the logistic function one obtains the partial credit model. For simplicity we assume in the following that $F(.)$ is the logistic function, although the derivations hold for more general distribution functions.

Various properties of the PCM have been used to obtain an interpretation of the parameters.  For example, it is easily seen that for $\theta_p=\delta_{ir}$ the probabilities of adjacent categories are equal, that is, $P(Y_{pi}=r)={P(Y_{pi}=r-1)}$. Moreover, response curve modes are always ordered. However, properties of this type seem not sufficient to obtain a satisfying interpretation of parameters.

While the cumulative model is derived as a coarser version of an underlying latent trait and the sequential model as a model for consecutive steps  an underlying process model for the PCM seems harder to obtain. Possible models will be considered later. First we consider the binary variables contained in the model.

\subsection{Binary Responses in the Partial Credit Model}\label{BininPCMP}

Consider the conditional binary variables defined by
\begin{equation}\label{eq:binaryPCMNew}
Y_{pi}^{(r)}=\left\{
\begin{array}{ll}
1&Y_{pi} \geq r \text{ given } Y_{pi} \in\{r-1, r\}\\
0&Y_{pi} < r \text{ given } Y_{pi} \in\{r-1, r\}
\end{array}
  \right. 
\end{equation}

It is easily seen that if the PCM holds  one has for the binary variables   
\begin{equation}\label{eq:PCM}
{P(Y_{pi}^{(r)}=1|Y_{pi} \in \{r-1,r\})}= F(\theta_p-\delta_{ir}),\quad r=1,\dots,k.
\end{equation}
Therefore, the   variables defined in (\ref{eq:binaryPCMNew}) are the binary variables contained in the PCM, which are also given in Table \ref{tab:ORDINAL}. 
It is straightforward to show that one can also start from the variables defined in (\ref{eq:binaryPCMNew}) and derive from them that the PCM
holds. In this sense the binary variables can be considered the \textit{building blocks} of the PCM and can be used to interpret the parameters.

In contrast to the cumulative model 
the sequential and the adjacent categories model contain \textit{conditional} binary   models, in the case of the sequential model  the condition is   $Y \ge r-1$ , and  $Y_{pi} \in \{r-1,r\}$ in the adjacent categories model. For the sequential model the conditioning comes from the modelling of consecutive steps which yields a straightforward interpretation of the model and its parameters. The conditioning in the adjacent categories model is much harder to handle, it is less distinctly motivated and gave rise to severe misunderstandings of the model. 

In the early  papers on the PCM it has been misunderstood as a model for consecutive steps.  
\citet{Masters:82} explicitly considered the example of solving the item $\sqrt{9.0/0.3-5}=?$ mentioned previously and explained the consecutive steps. It was said, for example:

\medskip
\begin{addmargin}[0.5 cm]{-0.0 cm}
\small
The third step in item $i$, for example, is from level 2 to level 3. The difficulty of this third
step governs how likely it is that a person who has already reached level 2 will complete the
third step to level 3.
(\citet{Masters:82}, p. 157)\\
\end{addmargin}

It seems that nowadays it is widely accepted that the partial credit models is not a model for consecutive steps. The model for consecutive steps is the sequential models, which is a distinctly different model, for further discussion see, for example, \citet{Tutz:90b},  \citet{verhelst1997steps}, \citet{verhelst2008some}, \citet{andrich2015problem}.
In later presentations the conditional structure of the PCM has been acknowledged.  \citet{masters2016partial} states correctly that the local comparison is the heart of the PCM, one conditions on a pair of adjacent categories and so eliminates all other response probabilities from consideration.

One consequence of the conditional definition of the binary variables is that they are not observable. If, for example, one observes $Y_{pi}=5$, the outcome of $Y_{pi}^{(1)}$ will not be observed since it postulates the condition $Y_{pi} \in \{0,1\}$. 
Nevertheless, one can try to investigate the structure of the potential outcomes of the sequence $(Y_{pi}^{(1)}\dots,Y_{pi}^{(k)})$. If $Y_{pi}$ takes value $r$, one knows which values the binary variables $Y_{pi}^{(r)}$ and $Y_{pi}^{(r+1)}$ take. 
The former distinguishes  between $\{r-1,r\}$ and the latter between $\{r,r+1\}$. If $Y_{pi}=r$ the binary variables have to be  $Y_{pi}^{(r)}=1$ and $Y_{pi}^{(r+1)}=0$. That means the sequence $(Y_{pi}^{(r)},Y_{pi}^{(r+1)})$ has to be $(1,0)$. 
Thus, adjacent binary variables yield outcomes of the form $(1,0)$. 

For the interpretation of parameters the conditional representations (\ref{eq:PCMDef}) and (\ref{eq:PCM}) are crucial. Since \textit{locally} (given categories $\{r-1,r\}$) a Rasch model holds one might use the interpretation of parameters that is familiar from the Rasch model. In an achievement test $\theta_p$ represents the ability, and $\delta_{ir}$ is the difficulty of observing $r$ rather than $r-1$ \textit{given the alternatives $\{r-1,r\}$}. What makes the interpretation less satisfying than for cumulative and sequential models  is that one cannot drop the conditioning. Therefore, one has also to be cautious when trying to interpret the binary models as models for the transition from category $r-1$ to category $r$ determined by the threshold $\delta_{ir}$. They might be seen as transition models, but again only locally. 
The representations (\ref{eq:PCMDef}) and (\ref{eq:PCM}) show only local properties  although changing one of the thresholds changes all of the probabilities. The response in any category is a function of all the thresholds. Alternative motivations and derivations of the PCM that have been given  are considered in the following.  


\subsection{Alternative Derivations of the Partial Credit model}\label{sec:BuildBlocks}


The local nature of the PCM makes it hard to find a model for the process by which a person reaches category. Therefore, in their   study of invariance properties of various ordinal models  \citet{peyhardi2014new} even conclude that "`adjacent models cannot be interpreted using latent variables"'. However, this conclusion lacks substance. It  is not hard to find a process that determines the total response probabilities.
It is tempting to imagine an underlying response process as a two-step model. In the first step a person chooses a pair of adjacent categories, in the second step he/she chooses one of the two categories. Since there are $k$ pairs of adjacent categories $(0,1),\dots,(k-1,k)$ in the first step one needs a probability function that chooses among the $k$ pairs.  In the second step, let the model for choosing $r$ rather than $r-1$, given $\{r-1,r\}$ were chosen in the initial step, be the local Rasch model (\ref{eq:PCMDef}). Since one considers two steps one might suspect that the distribution in the first step could be chosen freely.  However, the problem is that the distribution on the adjacent categories $(0,1),\dots,(k-1,k)$ in the first step is already fixed if one assumes the Rasch model in the second step. Therefore, person and item parameters already determine the probabilities in the first step and the process offers no additional insights.

In the following alternative and more potent derivations of the PCM are considered and properties of the involved binary variables are investigated.    

\subsubsection*{Deriving the Partial Credit Model from Independent and Guttman Variables}

\citet{andrich2013expanded} gave an expanded derivation of the threshold structure of the polytomous Rasch model, which was already considered in  \citet{andrich1978rating}. In  his article he strongly criticizes the findings of \citet{adams2012rasch} as misunderstandings centered around the so-called "`threshold disorder controversy"'. In doing so he considers various response spaces, among them the Guttman space, which is of special interest here because it provides an alternative derivation of the PCM.  In the following we first sketch the basic results. 

Let us consider $k$ \textit{independent} Bernoulli variables, $\tilde Y_{pi}^{(r)}$, $r=1,\dots,k$, $\tilde Y_{pi}^{(r)} \in \{0,1\}$, which follow a Rasch model 
\begin{equation}\label{eq:Label2}
P(\tilde Y_{pi}^{(r)}=1)= F(\theta_p-\delta_{ir}),\quad r=1,\dots,k.
\end{equation}
However, instead of observing  $\tilde Y_{pi}^{(1)},\dots,\tilde Y_{pi}^{(k)}$ one performs the following  \textit{Guttman space experiment}:
\begin{itemize}
\item[] (1) Observe the realizations of the independent Bernoulli variables $\tilde Y_{pi}^{(1)},\dots,\tilde Y_{pi}^{(k)}$.
\item[] (2) If for any $r<s$ one observes $\tilde Y_{pi}^{(r)}=0$ and $\tilde Y_{pi}^{(s)}=1$ the sequence of observations $\tilde Y_{pi}^{(1)},\dots,\tilde Y_{pi}^{(k)}$ is considered as not valid.
\end{itemize}
The effect is that one does not observe the $2^k$ possible vectors $(\tilde y_{pi}^{(1)},\dots,\tilde y_{pi}^{(k)})$,
$\tilde y_{pi}^{(r)} \in \{0,1\}$, but a reduced set of observations in the \textit{Guttman space}, which  is given by 
$\Omega^G=\{(y_1,\dots,y_k)|y_r \in \{0,1\}, y_r \ge y_s \text{ for all } r \le s\}$.
It contains all vectors that have the form $(1,\dots,1,0,\dots,0)$. Thus if one variable takes value zero the following variables also have to take the value zero. It is essential to distinguish between the independent Bernoulli variables and the  
variables that yield observations in the Guttman space. Let 
$Y_{pi}^{(1)},\dots,Y_{pi}^{(k)}$ denote the binary variables of the Guttman space, which are distinctly different from the independent  random variables $\tilde Y_{pi}^{(1)},\dots,\tilde Y_{pi}^{(k)}$. The latter are   the sequence of independent variables used to construct the response space. The former are the random variables representing the outcome in the Guttman space (after deleting the non valid outcomes of the original independent variables).  The Guttman variables are obtained by conditioning on $\tilde Y_{pi}^{(r)} \ge \tilde Y_{pi}^{(r+1)}$ for all $r$.

It is important that the definition of the Guttman space  used here postulates only $Y_r \ge Y_s$  for all  $r \le s$ (and $Y_r \in \{01\}$). It is \textit{not} assumed that the $Y_1,\dots,Y_k$ are a sequence of successes and failures, for which one assumes an ordering of the thresholds of the Rasch models that determine single responses.  One of the problems with Andrich's derivations is that when he uses the term Guttman space he  implies an ordering of thresholds, which he calls required ordering. This ``intended ordering of thresholds'' is not needed for the derivations in the following and might be a source of confusion. Ordering of thresholds might be a consequence when assuming a Guttman space, as has been shown in the derivation of the cumulative model, but is not needed for the definition of the space. A Guttman space as used here is a space of responses that has to be distinguished from a Guttman scale obtained for single items. We use the term Guttman scale only if  an ordering of thresholds is postulated. 

One obtains the following proposition, which  has also be given by \citet{andrich2013expanded}, although in slightly different form. A proof is given in the Appendix.

\begin{theorem}\label{th:BerGut}
If one defines the random variable $Y_{pi}=\sum_{s=1}^k  Y_{pi}^{(r)}$, which is the sum of the Bernoullis in the Guttman space,  one obtains for $Y_{pi}$ the PCM. 
\end{theorem}

The result is remarkable in several ways. First, the PCM is derived   from the assumption of binary variables.   The binary variables are defined without referring to a random variable with $k$ outcomes. Instead the variable $Y_{pi}$ is defined by the (sum of) binary variables. 
Second, one obtains the PCM from underlying \textit{independent} binary variables, which seems surprising. 
However, one should not forget that the  PCM is obtained by postulating the restriction to the Guttman space with $Y_{pi}$ representing the numbers of outcomes '1's that have been observed as a sequence of '1's not disturbed by zeros.
Without this restriction one would not obtain the PCM.


If $Y_{pi}=r$, that is, if one has the sequence $( Y_{pi}^{(1)},\dots, Y_{pi}^{(k)})=(1,\dots,1,0,\dots,0)$, with the last 1 occurring in the $r$-th component it is tempting to interpret it as a sequence of $r$ successful steps (the ones) followed by a sequence of $k-r$ unsuccessful steps (the zeros). However, one should be aware that one considers the sequence in the Guttman space, not in the space used for the construction, which is $(\tilde Y_{pi}^{(1)},\dots,\tilde Y_{pi}^{(k)})$. It is slightly irritating that \citet{andrich2013expanded}, when defining the sum of Bernoullis, does not clearly distinguish between these different random variables (see equ (5) and (8) in \citet{andrich2013expanded}). 
Although the sum $Y_{pi}=\sum_{s=1}^k  Y_{pi}^{(r)}$ defined in Proposition \ref{th:BerGut} takes  the same values as the random variable 
$\tilde Y_{pi}=\sum_{s=1}^k \tilde Y_{pi}^{(r)}$, namely $0,\dots,k$, the distributions of $Y_{pi}$ and $\tilde Y_{pi}$ differ, and consequently $Y_{pi} \ne \tilde Y_{pi}$.
Confusing $Y_{pi}$ and $\tilde Y_{pi}$ is dangerous because  $\tilde Y_{pi}$ is built from  dependent Bernoullis. If one does not distingiush between these sums one might try to interpret the response from the properties of the underlying experiment with independent observations.

\begin{table}[h!]
\caption{Contingency tables for the independent binary variables $(\tilde Y_{pi}^{(1)}, \tilde Y_{pi}^{(2)} )$ (left) and the  binary variables $( Y_{pi}^{(1)},  Y_{pi}^{(2)} )$ derived from conditioning }\label{tab:MARG}
\centering
 
  \begin{minipage}{0.4\textwidth}
    \begin{tabular}{cc|cc|c}
                           & \multicolumn{1}{c}{}            & \multicolumn{2}{c}{$\tilde Y_{pi}^{(2)}$}&                 \\
                           & \multicolumn{1}{c}{} & 1 & \multicolumn{1}{c}{0}&         \\ \cline{3-4}
                           & 1                        & $\tilde \pi_{11}$  & $\tilde \pi_{10}$                           & $\tilde \pi_{1+}$      \\
\raisebox{1.5ex}[-1.5ex]{$\tilde Y_{pi}^{(1)}$} & 0                        &
$\tilde \pi_{01}$ & $\tilde \pi_{00}$                           & $\tilde \pi_{0+}$      \\
\cline{3-4}
                           &  \multicolumn{1}{c}{}         & $\tilde \pi_{+1}$       & \multicolumn{1}{c}{$\tilde \pi_{+0}$}                          & 1      \\
\end{tabular}
\normalsize
  \end{minipage}
  \begin{minipage}{0.5\textwidth}
\begin{tabular}{cc|cc|c}
                           & \multicolumn{1}{c}{}            & \multicolumn{2}{c}{$Y_{pi}^{(2)}$}&                 \\
                           & \multicolumn{1}{c}{} & 1 & \multicolumn{1}{c}{0}&         \\ \cline{3-4}
                           & 1                        & $\pi_{11}$  & $\pi_{10}$                          & $\pi_{1+}$      \\
\raisebox{1.5ex}[-1.5ex]{$Y_{pi}^{(1)}$} & 0                        &
$-$ & $\pi_{00}$ & $\pi_{0+}$      \\ \cline{3-4}
                           &  \multicolumn{1}{c}{}         & $\pi_{+1}$      & \multicolumn{1}{c}{$\pi_{+0}$}                          & $n$      \\
\end{tabular}
\normalsize
  \end{minipage}
\end{table}

It is instructive  to investigate the distributions of the random variables implicitly defined by the Guttman experiment, which are far from being 
independent. 
Let us first consider the case of three categories with two independent binary variables $\tilde Y_{pi}^{(1)}, \tilde Y_{pi}^{(2)}$. 
The left panel in Table \ref{tab:MARG} shows the probabilities in a $(2\times 2)$-contingency table.
The marginal distributions of $\tilde Y_{pi}^{(r)}$  are specified by a binary Rasch model, 
\[
\tilde\pi_{1+}=P(\tilde Y_{pi}^{(1)}=1)= F(\theta_p-\delta_{i1}), \quad \tilde\pi_{+1}=P(\tilde Y_{pi}^{(2)}=1)= F(\theta_p-\delta_{i2}),
\]
or, equivalently,
\[
\frac{P(\tilde Y_{pi}^{(1)}=1)}{ P(\tilde Y_{pi}^{(1)}=0)}= \exp(\theta_p-\delta_{i1}), \quad \frac{P(\tilde Y_{pi}^{(2)}=1)}{P(\tilde Y_{pi}^{(2))}=0}= \exp(\theta_p-\delta_{i2}).
\]
According to the assumption of independence the probabilities in the table are given by 
$\tilde\pi_{ij}=\tilde\pi_{i+}\tilde\pi_{+j}$, $i, j \in \{1,2\}$.
 
The conditioning on $(\tilde Y_{pi}^{(1)}, \tilde Y_{pi}^{(2)} ) \in \{ (1,1), (1,0), (0,1) \}$ yields a  distribution of $(Y_{pi}^{(1)},Y_{pi}^{(2)})$ that is quite different from the distribution of  $ (\tilde Y_{pi}^{(1)}, \tilde Y_{pi}^{(2)})$. In the right panel of Table \ref{tab:MARG} the probabilities $\pi_{rs}=P((Y_{pi}^{(1)},Y_{pi}^{(2)})=(r,s))$ are shown. One cell is empty, the probabilities of the other cells are given by 
\[
\pi_{11}= \tilde \pi_{11}/s_{\pi}, \quad \pi_{10}= \tilde \pi_{10}/s_{\pi},\quad \pi_{00}= \tilde \pi_{00}/s_{\pi}, 
\]
where $s_{\pi}= \tilde \pi_{11}+\tilde \pi_{10}+\tilde \pi_{00}$. It is immediately seen that  $Y_{pi}^{(1)}$ and $Y_{pi}^{(2)}$ are not independent.

One obtains for the marginals $\pi_{r+}=P(Y_{pi}^{(1)}=r)$, $\pi_{+r}=P(Y_{pi}^{(2)}=r)$, which for simplicity are represented by the odds,  
\[
\frac{P( Y_{pi}^{(1)}=1)}{P( Y_{pi}^{(1)}=0)}= \frac{ \pi_{1+}}{\pi_{0+}}=\frac{\tilde\pi_{1+}}{\tilde\pi_{00}}=\frac{\tilde\pi_{1+}}{\tilde\pi_{0+}}\frac{1}{\tilde\pi_{+0}}={\exp(\theta_p-\delta_{i1})}\frac{1}{P(\tilde Y_{pi}^{(2)}=0)}
\]
\[
\frac{P( Y_{pi}^{(2)}=1)}{P( Y_{pi}^{(2)}=0)}= \frac{ \pi_{+1}}{\pi_{+0}}=\frac{\tilde\pi_{11}}{\tilde\pi_{+0}}= {\tilde\pi_{1+}}\frac{\tilde\pi_{+1}}{\tilde\pi_{+0}}=P(\tilde Y_{pi}^{(1)}=1){\exp(\theta_p-\delta_{i2})}
\]
Therefore, the marginal distribution of $Y_{pi}^{(1)}$ depends on $\tilde Y_{pi}^{(2)}$ and the marginal distribution of $Y_{pi}^{(2)}$ depends on $\tilde Y_{pi}^{(1)}$. The odds for $Y_{pi}^{(1)}$ are the odds of the Rasch model for $Y_{pi}^{(1)}$ modified by $P(\tilde Y_{pi}^{(2)}=0)$. Accordingly the odds for $Y_{pi}^{(2)}$ are the odds of the Rasch model for $Y_{pi}^{(2)}$ modified by $P(\tilde Y_{pi}^{(1)}=1)$.

However, it is interesting to consider the conditional distributions of  the variables $Y_{pi}^{(1)},Y_{pi}^{(2)}$. One  can  derive the following proposition (a generalization is given later).

\begin{theorem} \label{try} 
For $(Y_{pi}^{(1)},Y_{pi}^{(2)})$ one obtains the Rasch-type conditional distributions 
\begin{equation}\label{Threecat}
P( Y_{pi}^{(2)}=1|Y_{pi}^{(1)}=1)=F(\theta_p-\delta_{i2}),
\quad
P( Y_{pi}^{(1)}=1|Y_{pi}^{(2)}=0)=F(\theta_p-\delta_{i1}).
\end{equation} 
\end{theorem}

That means that  specific conditional distributions follow a Rasch model while the marginals of $Y_{pi}^{(1)}, Y_{pi}^{(2)}$ do not.  
It is easily derived how the binary variables are linked to the ordinal response in a PCM. Let us consider the first equation. The sum $Y_{pi}=Y_{pi}^{(1)}+   Y_{pi}^{(2)}$ takes values from $\{0,1,2\}$. Under the condition $Y_{pi}^{(1)}=1$ the variable $Y_{pi}$ is restricted to values 
$\{1,2\}$, and $Y_{pi}=2$ is equivalent to $Y_{pi}^{(2)}=1$. One obtains $P( Y_{pi}^{(2)}=1|Y_{pi}^{(1)}=1)=P(Y_{pi}=2|Y_{pi}\in $\{1,2\}$)$.
Since a similar transformation holds for  the second equation,  it is seen 
that the variables $Y_{pi}^{(1)}, Y_{pi}^{(2)}$ are just the conditional variables obtained as the binary variables contained in the PCM as derived in Section \ref{BininPCMP} (equ. (\ref{eq:binaryPCMNew})). They represent the Rasch models with local interpretation.

Proposition \ref{try} is useful to clarify the difference in the Guttman space that is used here and the Guttman spcace that was used   in the derivation  of the cumulative model from binary variables. For the  derivation of  the cumulative model in Section \ref{sec:BinBuild} it was assumed that the  \textit{marginals} follow binary Rasch models. Here, the binary Rasch models are found as  conditional models, not for the marginals, although they were assumed for the marginals of the underlying independent variables. Nevertheless, in both derivations, for the cumulative model and the PCM, the Guttman structure of the binary responses is postulated. 


\begin{table}[h!]
\caption{Operational Definition for Judging Essays With Respect to the Property of Setting}\label{tab:Judge}

\centering
\begin{tabularsmall}{@{}ccc@{}}
\toprule \multicolumn{3}{c}{}\\
    Fail(F)                & Pass (P) & Distinction (D)\\\midrule
Inadequate          &Discrete &Integrated and manipulated   \\

\bottomrule
\end{tabularsmall}
\end{table}

For illustration we consider a concrete example with only three ordered categories $0,1,2$ for the total response. As \citet{andrich2013expanded} we use a specific example from Harris (1991) in the assessment of essay writing with respect to the criterion of setting. Table \ref{tab:Judge} shows the categories Fail, Pass, and Distinction in the assessment of proficiency, which are obviously ordered categories.

One might try to interpret the components of the random vector $(Y_{pi}^{(1)},Y_{pi}^{(2)})$  as steps or transitions from one level to the next. They seem more suitable for an interpretation as successive transitions than the variables $(\tilde Y_{pi}^{(1)},\tilde Y_{pi}^{(2)})$ because for them $Y_{pi}^{(1)} \ge Y_{pi}^{(2)}$ holds. In the example considered here  one has two transitions. The first  transition, $Y_{pi}^{(1)}$, is from Fail to Pass (level 0 to 1), the second transition, $Y_{pi}^{(2)}$, is from  Pass to Distinction (level 1 to 2). If one accepts the interpretation of transitions, the conditional probabilities in the proposition could be interpreted as follows.
If a person succeeds in the first transition from the lowest to the medium level ($Y_{pi}^{(1)}=1$)
the conditional probability of succeeding in the second transition follows a Rasch model.
If a person fails in the second transition from the medium to the highest level ($Y_{pi}^{(2)}=1$)
the conditional probability of succeeding in the first transition follows a Rasch model.

It has to be emphasized that the transitions are definitely not steps that are tried in a consecutive order,  there is no ordering in the sense that $Y_{pi}^{(1)}$ is performed before $Y_{pi}^{(2)}$. 
It is crucial to understand that the variables are strongly associated but cannot be considered as consecutive variables as the Markov process in the sequential model. They have to be considered as strongly linked "simultaneous" transitions. In a Markov process one typically considers the conditional distribution of random variables given the past. Here also the conditional probability $P( Y_{pi}^{(1)}=1|Y_{pi}^{(2)}=0)$, which is a conditioning on the second transition, is relevant.  
It should also be noted that the Rasch model that holds for $P( Y_{pi}^{(1)}=1|Y_{pi}^{(2)}=0)$ does not mean that the second transition is determined by a Rasch model \textit{given the person is unable to perform the second transition}. The condition is \textit{given $Y_{pi}^{(2)}=0$}, that is, that the second transition is not successful.  


\subsubsection*{The General Case}

The properties of the PCM given so far are also found (in a slightly different form) in \citet{andrich2013expanded}. 
This is not the case for the results derived in the following. Let us start with  the generalization of Proposition \ref{try}.
Let the marginals of the independent variables given in (\ref{eq:Label2}) be denoted by 
\[
\tilde \pi_r=P(\tilde Y_{pi}^{(r)}=1)= F(\theta_p-\delta_{ir}),\quad r=1,\dots,k.
\]
One obtains the following result (for a proof, see Appendix).

\begin{theorem} \label{th:gen} 
For the binary variables $Y_{pi}^{(1)},\dots,Y_{pi}^{(k)}$ of the corresponding Guttman space one obtains  the Rasch-type conditional distributions 
\begin{equation} \label{equ:gen1}
P( Y_{pi}^{(r)}=1|Y_{pi}^{(r-1)}=1,Y_{pi}^{(r+1)}=0)= F(\theta_p-\delta_{ir})
\end{equation}
\end{theorem}

\medskip
The Rasch model is found for conditional probabilities of the event $Y_{pi}^{(r)}=1$ given the ``previous'' response $Y_{pi}^{(r-1)}$ takes value 1 \textit{and} the ``later'' response  $Y_{pi}^{(r+1)}$ takes value 0. It shows that the Rasch model holds \textit{locally conditional on the responses in adjacent categories}. 
As special cases one obtains for the extreme categories 
\begin{equation} \label{equ:gen2}
P( Y_{pi}^{(k)}=1|Y_{pi}^{(k-1)}=1)=F(\theta_p-\delta_{ik}), \quad P( Y_{pi}^{(1)}=1|Y_{pi}^{(2)}=0)=F(\theta_p-\delta_{i1}).
\end{equation}
Thus, for the extreme Guttman variables $Y_{pi}^{(1)}$ and $Y_{pi}^{(k)}$, the conditioning is on 
just one event,  a ``previous'' or ``later'' one. For $k=2$ these are the conditional probabilities given in Proposition \ref{try}.

Let us briefly summarize the given derivations. We started with assuming 
the existence of independent Bernoullis $\tilde Y_{pi}^{(1)},\dots,\tilde Y_{pi}^{(k)}$ that marginally follow a Rasch model. 
Then a Guttman space experiment is performed that turns the original variables into Guttman variables $Y_{pi}^{(1)},\dots,Y_{pi}^{(k)}$.
An ordinal response variable is defined by $Y_{pi}=\sum_{s=1}^k  Y_{pi}^{(r)}$ and it was shown that for this response variable the PCM holds.

\citet{andrich2013expanded} tries to run the argument in the reverse direction. He starts with an ordinal response variable, for which the PCM holds, and tries to infer the existence of a \textit{latent} Guttman space, which is linked to independent Bernoullis. Before considering his derivations in more detail we investigate the Guttman space (or, equivalently, the Guttman variables) in the PCM. In particular, it will be shown that the Guttman variables that are linked to the PCM are much easier to obtain than in Andrich's derivation.   


\subsection*{The Guttman Space in the Partial Credit Model}\label{sec:ExGut}

Let us first assume that for   $Y_{pi} \in \{0,\dots,k\}$ the PCM holds. Then one can \textit{define}  Guttman variables by
\begin{equation}\label{eq:Defbin}
   (Y_{pi}^{(1)}=1,\dots,Y_{pi}^{(r)}=1,Y_{pi}^{(r+1)}=0,\dots,Y_{pi}^{(k)}=0) \text{ if }    Y_{pi}=r.
\end{equation}\ 
It should be noted that the definition works   since only  outcomes of the form $(1,\dots,1,0,\dots,0)$ are allowed. Only then the number of possible outcomes in the Guttman space is equal to the number of response categories of $Y_{pi}$ and the probability for all vectors from the Guttman space is determined by the probabilities $P(Y_{pi}=r)$, and therefore the variables $Y_{pi}^{(r)}$ are well defined. It is noteworthy that in the definition of the Guttman vectors no conditioning is needed. In Table \ref{tab:Three} the correspondence between $Y_{pi}$ and the binary variables is given for the case of three response categories.
\begin{table}[h!]
\caption{Guttman variables $Y_{pi}^{(1)},Y_{pi}^{(2)}$ defined by the ordinal response $Y_{pi} \in \{0,1,2\}$}\label{tab:Three}

\centering
\begin{tabularsmall}{@{}ccc@{}}
\toprule \multicolumn{2}{c}{}\\
    $Y_{pi}$                & $(Y_{pi}^{(1)},Y_{pi}^{(2)})$   \\\midrule
0          &(0,0)    \\
1          &(1,0)    \\
2          &(1,1)    \\
\bottomrule
\end{tabularsmall}
\end{table}

From the definition of the binary variables one obtains ($k=2$)
\begin{align*}
   P(Y_{pi}=1|Y_{pi} \in \{0,1\})&= P((Y _{pi}^{(1)},Y_{pi}^{(2)})=(1,0)|(Y_{pi}^{(1)},Y_{pi}^{(2)}) \in \{(0,0),(1,0)\})\\
  &= P(Y_{pi}^{(1)}=1|Y_{pi}^{(2)}=0),\\
 P(Y_{pi}=2|Y_{pi} \in \{1,2\})&= P((Y _{pi}^{(1)},Y_{pi}^{(2)})=(1,1)|(Y_{pi}^{(1)},Y_{pi}^{(2)}) \in \{(1,0),(1,1)\})\\
  &= P(Y_{pi}^{(2)}=1|Y_{pi}^{(1)}=1). 
\end{align*}
Since the PCM is assumed to hold one has
\[
P( Y_{pi}^{(2)}=1|Y_{pi}^{(1)}=1)=F(\theta_p-\delta_{i2}),
\quad
P( Y_{pi}^{(1)}=1|Y_{pi}^{(2)}=0)=F(\theta_p-\delta_{i1}),
\]
which are the equ. (\ref{Threecat}) from Proposition \ref{try}. As shown in the Appendix  one obtains in the general case:

\begin{theorem} \label{th:genobsGut} 
Let the PCM hold and $Y_{pi}^{( 1)},\dots,\dots,Y_{pi}^{(k)}$ be the Guttman  variables defined by (\ref{eq:Defbin}). Then, one has 
\begin{equation} \label{equ:gen11}
P( Y_{pi}^{(r)}=1|Y_{pi}^{(r-1)}=1,Y_{pi}^{(r+1)}=0)= F(\theta_p-\delta_{ir}).
\end{equation}
\end{theorem}

The condition (\ref{equ:gen11}) is the same as obtained when assuming independent Bernoullis and then conditioning on the Guttman space (equ. (\ref{equ:gen1})).
Therefore, when assuming that the PCM holds one can construct a   Guttman space, or equivalently, define Guttman variables, for which  condition (\ref{equ:gen11}) holds. 

It is crucial that also the reverse holds. One can start from   Guttman variables, for which  condition (\ref{equ:gen11}) holds and derive that the PCM holds. This is seen from the following proposition, which can be easily derived.  

\begin{theorem} \label{th:gen3} 
Let $Y_{pi}^{(1)},\dots,Y_{pi}^{(k)}$ be Guttman variables, for which (\ref{equ:gen11}) holds. Then, for the sum $Y_{pi}= Y_{pi}^{(1)}+\dots+Y_{pi}^{(k)}$ the PCM holds.
\end{theorem}

In summary, the propositions \ref{th:genobsGut} and \ref{th:gen3}  yield the structure

\begin{addmargin}[0.5 cm]{-0.0 cm}
\small
\begin{align*}
&\text{The PCM holds}   &\Rightarrow    &\phantom{aaa}\text{Condition (12) holds} \\
&\   &   &\phantom{aaa}\text{  for the Guttman variables defined by (11)}\\
& \text{The PCM holds for the} &\Leftarrow    & \phantom{aaa}\text{  Guttman variables that fulfill }\\
&\text{sum of Guttman variables} &    & \phantom{aaa}\text{condition (12) exist}
\end{align*}
\end{addmargin}

Thus, it has been established that  the existence of an observable Guttman space, for which (\ref{equ:gen11}) holds,   is a sufficient and necessary condition for the PCM. In particular, when deriving the PCM the independent variables are not needed. It is sufficient to start from   Guttman variables for which the condition (\ref{equ:gen11}) holds.

Since the condition  is central to the Guttman space that is linked to the PCM let us consider its meaning more closely.
The conditioning on responses of the adjacent variables in (\ref{equ:gen11}) reflects the local nature of the PCM. 
It shows that a binary Rasch model holds for the preference of category $r$ over $r-1$ given $r-1$ is preferred over $r-2$ ($Y_{pi}^{(r-1)}=1$)
and $r+1$ is not preferred over $r$ ($Y_{pi}^{(r+1)}=0$). Thus preferences   are modeled locally given specific  preferences in adjacent categories are present or not. 
Condition (\ref{equ:gen11}) can be seen as a motivation for   an underlying process model. When a person compares all pairs of adjacent categories and decides that he/she  prefers $r-1$   over $r-2$  but definitely does not prefer  $r+1$ over $r$ the tendency to choose
$r$ among the pair  $\{r-1,r\}$ is determined by the binary Rasch model. However, in the local modeling actually four categories are involved, $r-2,r-1,r,r+1$.

\subsection*{Alternative Derivations of the  Guttman Space from the Partial Credit Model}\label{sec:ExGut}

In the previous sections it has in particularly  been shown how the PCM can be derived from the assumption of independent Bernoullis and conditioning to obtain Guttman variables. 
\citet{andrich2013expanded} also investigates the reverse derivation, he aims at ``establishing the Guttman response space from the structure of the PCM'' and tries to show that ``the Guttman structure with a hypothesized threshold order  is a necessary and sufficient condition for the PCM'' (p.99). 

Let us consider first the aim to establish  the Guttman response space from the structure of the PCM. \citet{andrich2013expanded} uses the following proposition that seems useful (see also \citet{luo2005relationship}):

\begin{theorem} \label{th:Andrich} 
Let $\tilde \pi_r= F(\theta_p-\delta_{ir}), r=1,\dots,k$.  The probabilities of the PCM can be represented as
\begin{equation} \label{equ:gen12}
P( Y_{pi}=r)= \tilde \pi_1\dots\pi_r(1-\tilde\pi_{r+1})\dots(1-\tilde\pi_{k})/\tilde s_{\pi},
\end{equation}
where $\tilde s_{\pi}=\sum_{j=1}^k \prod_{s=1}^j \tilde \pi_j\prod_{s=j+1}^k (1-\tilde \pi_j)$.
\end{theorem}

The proposition indeed suggests that there is a Guttman space behind the PCM. However, as has been  shown in the previous section the Guttman variables can be defined directly. The indirect way to infer the Guttman structure, which is briefly considered in the following, is much more difficult to understand.


In his derivation of a Guttman space, \citet{andrich2013expanded}  starts with a random variable $X_{pi}\in \{0,\dots,k\}$ and lets  
$P(X_{pi}=r|\Omega^{'})$ denote the probability of the response $r$ in the space $\Omega^{'}$. The space $\Omega^{'}$  is deliberately left undefined. It cannot be meant to be the space of the realizations of $X_{pi}$ because this space is trivially the set $\{0,\dots,k\}$. However, one can assume that it is the underlying probability space from which the random variable  is obtained as a mapping $X_{pi}:{\Omega^{'}} \rightarrow \{0,\dots,k\}$.   In the next step he "defines"  latent, dichotomous variables $Z_{pi}^{(r)}\in \{0,1\}$ by 
\begin{equation}\label{eq:And}
P(Z_{pi}^{(r)}=1|\Omega^{'}_{r-1,r})=\frac{P(Z_{pi}=r|\Omega^{'})}{P(Z_{pi}=r-1|\Omega^{'})+P(Z_{pi}=r|\Omega^{'})}=F(\theta_p-\delta_{ir}).
\end{equation}
A space $\Omega^{'}_{r-1,r}$ is introduced but not defined. Only the subscript signals that it must be linked to the values $r-1,r$. Thus the conditioning can only mean conditioning on $X_{pi} \in \{r-1,r\}$ or, equivalently $\{\omega: X_{pi}(\omega)\in \{r-1,r\} \}$. 
\citet{andrich2013expanded} lets the $\tilde \pi_r$ in Proposition \ref{th:Andrich} be defined by $\tilde \pi_r=P(Z_{pi}^{(r)}=1|\Omega^{'}_{r-1,r})$ and infers that   $P( Y_{pi}=r|\Omega^{'})= \tilde \pi_1\dots\pi_r(1-\tilde\pi_{r+1})\dots(1-\tilde\pi_{k})/\tilde s_{\pi}$ holds. Although this is not needed to obtain the result in Proposition \ref{th:Andrich}, it is used to infer  the space $\Omega^{'}$. It turns out that 
$\Omega^{'}$ is the Guttman space  and $\Omega^{'}_{r-1,r}$ is the Guttman space conditioning on responses in $\{r-1,r\}$.

The derivations seem unnecessarily complicated. The use of response spaces is not always clear and tends to confuse most readers unfamiliar with the concept of spaces. As has been shown above one can derive the Guttman variables directly, they can be defined from the other  random variables without reference to spaces.

\citet{andrich2013expanded} goes one step further, he wants to infer the existence of independent binary variables that yield the variables 
$Z_{pi}^{(1)}\dots,Z_{pi}^{(k)}$ as conditional responses, in his words to ``infer the existence of a hypothetical, complete space $\Omega$ of which $\Omega^G$ is a subspace''. However, given the PCM one may define the variables in the Guttman space, which has been done  above, but it seems hard to define independent response vectors $\tilde Y_{pi}^{(1)}\dots,\tilde Y_{pi}^{(k)}$, that yield  the variables 
$Z_{pi}^{(1)}\dots,Z_{pi}^{(k)}$ as conditional responses. Consequently, they were not explicitly defined in \citet{andrich2013expanded}. What can be done is to start from  independent variables with the right probabilities  and then derive conditional variables. This is exactly the derivation given  in Proposition \ref{th:BerGut}. It means that by assuming the right probabilities for the independent variables one obtains the Guttman variables as the conditional responses. It amounts to an embedding of the Guttman variables into a space of independent variables. But the starting point are the independent variables, their existence is not inferred by construction, it is just shown that one can postulate variables that are compatible with  the PCM.  From this embedding one can not infer that  the ``PCM with a hypothesized threshold order is a necessary and sufficient condition for the PCM''.


\citet{andrich2013expanded} concludes that the Guttman space $\Omega^{G}$ is manifest (observable) but the conditional space $\Omega^{G}_{r-1,r}$ is latent. Since spaces are used it is not obvious which random variable are actually considered. Therefore, let us again consider the 
condition $P( Y_{pi}^{(r)}=1|Y_{pi}^{(r-1)}=1,Y_{pi}^{(r+1)}=0)= F(\theta_p-\delta_{ir})$, which is the central condition in the Guttman space derived in Proposition \ref{th:gen3}. It has been shown that the condition can also be given in the form 
\begin{align*}
&P(Y_{pi} \ge r|Y_{pi} \in \{r-1,r\})=F (\theta_p-\delta_{ir}).  
\end{align*}
Therefore the condition is equivalent to postulating $P(\bar Y_{pi}^{(r)}=1|\bar Y_{pi}^{(r)}\in \{0,1\})=F(\theta_p-\delta_{ir})$, where  
\begin{equation}\label{eq:AndHelp}
\bar Y_{pi}^{(r)}=\left\{
\begin{array}{ll}
1&Y_{pi} \geq r \text{ given } Y_{pi} \in\{r-1, r\}\\
0&Y_{pi} < r \text{ given } Y_{pi} \in\{r-1, r\}
\end{array}
\right.
\end{equation}
The variables $\bar Y_{pi}^{(r)}$ are well defined but in contrast to $Y_{pi}$ they are \textit{not observable}. If one observes $Y_{pi}$ one knows that $\bar Y_{pi}^{(r)}$, which locally compares $r-1$and $r$, took value 1   and $\bar Y_{pi}^{(r+1)}$, which locally compares $r$and $r+1$, took value 0. Therefore, one obtains $(\bar Y_{pi}^{(r)},\bar Y_{pi}^{(r)})=(1,0)$. Since the condition varies across variables nothing is known about the value of other pairs of adjacent variables. It should be noted that the variables $\bar Y_{pi}^{(r)}$ are equivalent to the conditional variables defined in (\ref{eq:binaryPCMNew}). They seem to be the variables that determine the the conditional space $\Omega^{G}_{r-1,r}$ considered by Andrich. However, observable or not, they are not Guttman variables, because they always have the form $(\bar Y_{pi}^{(r)},\bar Y_{pi}^{(r)})=(1,0)$.

\blanco{
It can be shown that for the Guttman space equ. (\ref{equ:gen1}) from Proposition \ref{th:gen} holds. By conditioning on 
$Y_{pi} \in \{r-1,r,r+1 \}$, so that the variables are defined, one obtains

\begin{align*}
P( Y_{pi}^{(r+1)}=1|Y_{pi}^{(r)}&=1,Y_{pi} \in \{r-1,r,r+1\})=P( Y_{pi} \ge r+1|Y_{pi} \ge r,Y_{pi} \in \{r-1,r,r+1\})\\
&=P( Y_{pi} = r+1|Y_{pi} \in \{r,r+1\})=F(\theta_p-\delta_{i,r+1})
\end{align*}
}

Let us make some further remarks. First,  Guttman spaces are by no means an exclusive feature of the PCM. As has been shown in previous sections Guttman spaces occur also in derivations and representations of the cumulative and sequential models. Starting from these models one can always define Guttman type variables. Second, in none of the derivations concerning  the PCM it was assumed that thresholds are ordered, which is important since Andrich concludes that thresholds in the PCM have to be ordered.

\subsection{The PCM as a Model for Ordered Responses and the Ordering of Thresholds}

It is obvious that the cumulative and the sequential models by construction are ordered response models.
Also the PCM explicitly exploits that categories are ordered. This can be seen from the binary models contained in the model. If one considers the conditional response given categories $r-1, r$ one obtains the binary Rasch model
${P(Y_{pi}=r|Y_{pi} \in \{r-1,r\})}= \exp(\theta_p-\delta_{ir})/(1+\exp(\theta_p-\delta_{ir}))$. Thus, if the person parameter $\theta_p$ increases the probability of observing the higher category $r$ increases. Since the property holds for \textit{all} pairs of adjacent categories  definitely an ordering of categories is assumed and exploited by the model.

Moreover, the PCM  is stable under the reverse permutation but not under other permutations, which makes it an ordinal model.  
It should be noted that ordering of categories is used but  not in the sense of upward or downward ordering. More concrete, it can be shown that the model also holds for the reversed order  obtained from the reverse permutation $p(r)=k-r$, $r=0,\dots,k$. This property holds not only for the PCM but for all adjacent categories models with symmetric function $F(.)$, see \citet{peyhardi2014new} for a careful  analysis of invariance properties of ordered response models.  

Alternative and useful concepts of ordering were considered by \citet{adams2012rasch} and dismissed by \citet{andrich2013expanded}. \citet{andrich2013expanded} embeds his definition of order "in the intended relative difficulties of the thresholds in the resolved design". 
The resolved design is considered a "thought experiment", it "imagines judges making independent decisions". It corresponds to the independent variables $\tilde Y_{pi}^{(1)},\dots,\tilde Y_{pi}^{(k)}$ used to derive the PCM. 

By postulating "that the empirical ordering of categories is equivalent to the empirical ordering of successive thresholds that define adjacent categories on a continuum" (\citet{andrich2013expanded}, p 88), he links the order to the thresholds. This makes sense only because he argues that thresholds have to be ordered, which is just a postulate but nothing in the structure of the PCM demands ordering.    

The main problem with the derivations of \citet{andrich2013expanded} is the link between the so-called resolved
design and the Guttman space.  With reference to the three categories case he states
\medskip
\begin{addmargin}[0.5 cm]{-0.0 cm}
\small
Consider a design in which a single response in one of the three ordered categories
is resolved into two independent dichotomous decisions. The response
might be a self-report or made by a judge. For efficiency of exposition, we use the
latter case. Specifically, instead of one judge assigning each essay into one of three
categories, two different judges work independently, one judge decides if each essay
meets the standard at Pass, the other decides if the same essays meet the standard at
Distinction.  (\citet{andrich2013expanded}, p. 81)\\.
\end{addmargin}

In this "`resolved design"' the two independent decisions are linked to thresholds which are assumed to be ordered, which is referred to as 'intended category order'. In Andrich' words: 
\medskip
\begin{addmargin}[0.5 cm]{-0.0 cm}
\small
As a direct consequence of the intended category order, the a priori specification
is that the threshold at Distinction is more difficult than the threshold at Pass, ...
It is stressed that in this design, it is the continuum that is dichotomized at two
places (thresholds) with independent decisions made at each threshold.(\citet{andrich2013expanded}, p. 82)\\.
\end{addmargin}

When constructing response spaces one can indeed start  with independent binary variables, and, when interpreting them as independent dichotomous decisions of judges, one might postulate ordered thresholds. After obtaining Guttman variables by conditioning one obtains an ordered response variable by considering the sum of Guttman variables.  
For the dependent binary variables in the Guttman space one obtains ordered thresholds, since they have been postulated from the start.
The 'required order of thresholds' in the PCM is inherited from the order in the original independent binary variables.

If one starts from the PCM, that means, without restricting thresholds to be ordered, one can define Guttman variables as in Section \ref{sec:ExGut}, but the independent variables cannot be explicitly defined. Although one might find independent variables that generate the Guttman variables by conditioning this brings us back to starting from the independent variables. Thus, the claim "that inference can be made as if   responses at the thresholds were experimentally independent" (p. 80) is merely supported  by the fact that the PCM can be derived from independent variables. 

In the derivation of the PCM a crucial step is the transition from the independent variables to the Guttman variables.
There is no obvious reason why independent Bernoullis should generate only observations in the Guttman space, although it  is an   useful, constraint to obtain the PCM.  Actually, the order  the order of categories is introduced by restricting responses to the Guttman space. By assuming only responses of the form $(1,\dots,1,0,\dots,0)$ the order of categories is established, not by the ordering of thresholds as claimed by Andrich. Independent binary variables have no ordering, they can simply be permuted.

The use of the Guttman structure and Guttman space is not always clear in Andrich's paper. In our use  Guttman space means restriction to possible outcomes $(1,\dots,1,0\dots,0)$, without postulating ordering of any thresholds. 
When defining Guttman variables after starting from the PCM, one obtains a Guttman space in the latter sense. Andrich sometimes seems to use a different definition, when considering 'the Guttman structure, as a hypothesis of the threshold order' (p.93).  
He  emphasizes "that because a Guttman vector could not be chosen without the specification of threshold order, the inference depends on the a priori specification of threshold order in the resolved design, which in turn arises from the specified category order in the standard design " (p.89). However, starting from the PC without ordered thresholds yields Guttman type vectors without the specification of threshold order. 
He also concludes that 'the Guttman structure with a hypothesized threshold order is a necessary and sufficient condition for the PRM'(\citet{andrich2013expanded}, p. 99).  However, as shown in Section \ref{sec:Cum} the Guttman structure with a threshold order yields a cumulative model (if one assumes the Rasch model for the single Bernoullis and uses the sum as response). The PCM follows only if one assumes the Rasch model for underlying independent variables and then assumes that a Guttman experiment is performed.  It is not established that it is a necessary condition for the PCM because the PCM does not assume thresholds to be ordered.  


In summary, the PCM  can be derived from a Guttman space but without any threshold ordering involved. Thus the PCM does not assume that thresholds are ordered. At some points   Andrich seems to agree: 

\medskip
\begin{addmargin}[0.5 cm]{-0.0 cm}
\small
Indeed, there is nothing in the structure, nor in the estimation process, that forces the values
of thresholds estimates to be ordered. If there were such a feature, then the ordering
would be a property of the model (as in the graded response model), and not a property of the data, and therefore could not disclose problems with the category ordering. The consistent argument made previously, and repeated here, is for evidence
that the thresholds are ordered in the data, as disclosed by the PRM. (\citet{andrich2013expanded}, p. 103)\\.
\end{addmargin}

The statement is peculiar in itself. Data can be measured on an ordinal scale level, but data do not have thresholds, so how can ordering of thresholds be a property of the data. Thresholds are defined through the model and therefore are part of the modelling, not the data. 

As stated before the category ordering is not determined by the ordering of thresholds and therefore there are no problems to be disclosed with the category order. Nothing "goes awry" in the empirical ordering of the categories if the ordering of thresholds is violated. 
Of course, one has to distinguish between the "empirical ordering" ordering of thresholds and the structure of the model. The structure of the PCM does not impose an ordering of thresholds. Thus empirically, which can only mean by  estimation, one can investigate if (estimated) thresholds are ordered or not. 
However, a PCM with the additional assumption of ordered thresholds would be a submodel, that is, a more restrictive version  of the PCM. 
Thresholds with an inverse order would be an anomaly in the latter model but not in the PCM. 

\section{Concluding Remarks}
The role of binary variables in classical ordered response models has been investigated. In particular it has been shown that the Guttman structure is found in all three model types. Some caution is warranted if one considers the Guttman structure as a necessary and sufficient condition for a latent trait model since it depends on the definition of the ordered response arising from the Guttman structure.

If one has a Guttman structure and a Rasch model with ordered thresholds holds for the binary variables it can be inferred that the  
cumulative or graded response model holds for the sum of the Guttman variables. The reverse derivation starts from the cumulative model. If it holds one  indeed finds a Guttman structure for the   Bernoulli variables that use splits in the space of ordered categories. They follow a Rasch model with ordered thresholds  
 
If one has a Guttman structure and a Rasch model for the \textit{conditional distributions} $Y_{pi}^{(r)}|Y_{pi}^{(r-1)}=1$ it can be inferred that for the sum of the Guttman variables the sequential model holds. If one starts from the sequential model one  finds a Guttman structure for the   Bernoulli variables that use conditional splits in the space of ordered categories. They follow a Rasch model with  thresholds that are not necessarily ordered.  

If  one has a Guttman structure and a Rasch model for specific {conditional distributions} (given in (\ref{equ:gen11})) it can be inferred that for the sum of the Guttman variables the PCM holds. If one starts from the PCM one  finds a Guttman structure for the Bernoulli variables
defined in (\ref{eq:Defbin}), for them $Y_{pi}^{(r)}|Y_{pi}^{(r-1)}=1,Y_{pi}^{(r+1)}=0$ locally follows a Rasch model. 

We considered classical ordered latent trait models from the families of cumulative, sequential and adjacent categories models. More recently, alternative models have been proposed, in particular, tree-based models, which  were considered among others by \citet{de2012irtrees},  
\citet{bockenholt2013modeling}, \citet{khorramdel2014measuring},   \citet{bockenholt2016measuring} and \citet{bockenholt2017response}.
Tree-based models assume a nested structure with the building blocks given as binary models. For them the role of binary models is obvious 
since they are built on them.

\section*{Appendix}

\subsection{Proof Proposition \ref{th:BerGut}} 
In Proposition   \ref{th:BerGut} it is assumed that for the independent variables $\tilde Y_{pi}^{(r)}$ the marginals are given by Rasch models,
\[
\tilde \pi_r=P(\tilde Y_{pi}^{(r)}=1)= F(\theta_p-\delta_{ir}),\quad r=1,\dots,k.
\]
Let $s_{\pi}$ be defined  by
\[
s_{\pi}=\sum_{i_r \ge i_s \text{ if } r < s} P(\tilde Y_{pi}^{(1)}=i_1,\dots,\tilde Y_{pi}^{(1)}=i_k).
\]                       
It denotes the sum of probabilities for all cells that are in the corresponding Guttman space. 
Then one obtains for the sum of   Guttman vectors, which are defined by conditioning on the Guttman space (under the assumption  that none of the  probabilities are zero),
\begin{align*}
P(Y_{pi}&=r|Y_{pi} \in \{r-1,r\}) = P(Y_{pi}=r)/(P(Y_{pi}=r)+P(Y_{pi}=r-1))\\
&= \frac{\tilde \pi_1 \dots \tilde \pi_{r}(1-\tilde \pi_{r+1})\dots(1-\tilde \pi_{k})/s_{\pi}}
{(\tilde \pi_1 \dots \tilde \pi_r(1-\tilde \pi_{r+1})\dots(1-\tilde \pi_{k})+\tilde \pi_1 \dots \tilde \pi_{r-1}(1-\tilde \pi_{r})\dots(1-\tilde \pi_{k}))/s_{\pi}}\\
&= \frac{\tilde\pi_{r}(1-\tilde \pi_{r+1})\dots(1-\tilde \pi_{k}) }
{\tilde\pi_{r}(1-\tilde \pi_{r+1})\dots(1-\tilde \pi_{k}) +(1-\tilde \pi_{r})\dots(1-\tilde \pi_{k})}
= \frac{\tilde\pi_{r}}{\tilde\pi_{r}+(1-\tilde\pi_{r})}=\tilde\pi_{r}.
\end{align*}
Thus, one has   $P(Y_{pi} =r|Y_{pi} \in \{r-1,r\})=F(\theta_p-\delta_{ir})$, which means that for $Y_{pi}$ the PCM holds.

\subsection{Proof Proposition \ref{th:gen}} 
In Proposition   \ref{th:gen} it is assumed that for the independent variables $\tilde Y_{pi}^{(r)}$ the marginals are given by Rasch models,
\[
\tilde \pi_r=P(\tilde Y_{pi}^{(r)}=1)= F(\theta_p-\delta_{ir}),\quad r=1,\dots,k.
\]
Let $s_{\pi}$ again denote the sum of probabilities for all cells that are in the corresponding Guttman space, given by
\[
s_{\pi}=\sum_{i_r \ge i_s \text{ if } r < s} P(\tilde Y_{pi}^{(1)}=i_1,\dots,\tilde Y_{pi}^{(1)}=i_k)
\]                       
Then one obtains 

\begin{align*}
&P( Y_{pi}^{(r)}=1|Y_{pi}^{(r-1)}=1,Y_{pi}^{(r+1)}=0)= \frac{P(Y_{pi}^{(r-1)}=1,Y_{pi}^{(r)}=1,Y_{pi}^{(r+1)}=0)}{P(Y_{pi}^{(r-1)}=1,Y_{pi}^{(r+1)}=0)}\\
&=\frac{\tilde \pi_1 \dots \tilde \pi_{r}(1-\tilde \pi_{r+1})\dots(1-\tilde \pi_{k})/s_{\pi}}
{\{\tilde \pi_1 \dots \tilde \pi_{r}(1-\tilde \pi_{r+1})\dots(1-\tilde \pi_{k})/s_{\pi}+\tilde \pi_1 \dots \tilde \pi_{r}(1-\tilde \pi_{r})\dots(1-\tilde \pi_{k})\}/s_{\pi}}\\
&= \frac{\tilde\pi_{r}}{\tilde\pi_{r}+(1-\tilde\pi_{r})}=\tilde\pi_{r} = F(\theta_p-\delta_{ir}).
\end{align*}

\subsection{Proof Proposition \ref{th:genobsGut}}

Let the PCM hold and $Y_{pi}^{( 1)},\dots,\dots,Y_{pi}^{(k)}$ be the Guttman  variables defined by (\ref{eq:Defbin}). Then, one has 

\begin{align*}  
&P( Y_{pi}^{(r)}=1|Y_{pi}^{(r-1)}=1,Y_{pi}^{(r+1)}=0)
 =\frac{P(Y_{pi}^{(r-1)}=1,Y_{pi}^{(r)}=1,Y_{pi}^{(r+1)}=0)}{P(Y_{pi}^{(r-1)}=1,Y_{pi}^{(r+1)}=0)}\\
&=\frac{P(Y_{pi}=r)}{P(Y_{pi}=r)+P(Y_{pi}=r-1)} =F(\theta_p-\delta_{ir}). 
\end{align*}

\blanco{

\subsection{Proof Proposition \ref{th:gen3}}

Let $Y_{pi}^{( 1)},Y_{pi}^{(2)}$ be Guttman variables, for which (\ref{Threecat}) holds. With $\pi_{ij}=P(Y_{pi}^{(1)}=i,Y_{pi}^{(2)}=j)$ one obtains the equations
\begin{align*}
&F(\theta_p-\delta_{i1})=P( Y_{pi}^{(1)}=1|Y_{pi}^{(2)}=0)= \pi_{10}/(\pi_{10}+\pi_{00}),\\
&F(\theta_p-\delta_{i2})=P( Y_{pi}^{(2)}=1|Y_{pi}^{(1)}=1)=\pi_{11}/(\pi_{11}+\pi_{10}),\\
&\pi_{10}+\pi_{00}+\pi_{11}=1.
\end{align*}
Solving for $\pi_{10},\pi_{00},\pi_{11}$ yields immediately that the PCM holds.

Since the binary variables follow a Guttman structure  
$Y_{pi}=r$ is equivalent to 
\[
Y_{pi}^{(1)}=1,\dots, Y_{pi}^{(r-1)}=1,Y_{pi}^{(r)}=1,Y_{pi}^{(r+1)}=0,\dots,Y_{pi}^{(k)}=0, 
\] 
and 
$Y_{pi}=r-1$ is equivalent to
\[
Y_{pi}^{(1)}=1,\dots, Y_{pi}^{(r-1)}=1,Y_{pi}^{(r)}=0,Y_{pi}^{(r+1)}=0,\dots,Y_{pi}^{(k)}=0, 
\]
The two vectors differ only in the variable $Y_{pi}^{(r)}$. Therefore one obtains
\[
P(Y_{pi}=r|,Y_{pi} \in \{r-1,r\}) =P(Y_{pi}^{(r)}=1|)
\]
}

\bibliography{literatur}
\end{document}